\documentclass[11pt,a4paper]{article}
\usepackage[LGR,T1]{fontenc}
\usepackage[latin9]{inputenc}
\usepackage{units}
\usepackage{amsmath}
\usepackage{graphicx}
\usepackage{ulem}

\makeatletter

\pdfpageheight\paperheight
\pdfpagewidth\paperwidth

\ProvideTextCommand{\~}{LGR}[1]{\char126#1}

\pdfoutput=1
\usepackage{jcappub}

\title{\boldmath Gravitational Wave luminosity distance in viscous cosmological models}

\author[a]{Giuseppe Fanizza,}
\author[b]{Eliseo Pavone}
\author[b]{and Luigi Tedesco}

\affiliation[a]{Instituto de Astrofis\'ica e Ci\^encias do Espa\c{c}o,
Faculdade de Ci\^encias da Universidade de Lisboa,
Edificio C8, Campo Grande, P-1740-016, Lisbon, Portugal}
\affiliation[b]{Dipartimento di Fisica, Universit\'a di Bari, via Amendola 173, 70126 Bari, Italy and Istituto Nazionale di Fisica Nucleare, Sezione di Bari, Italy }

\emailAdd{g.fanizza@fc.ul.pt}
\emailAdd{e.pavone2@studenti.uniba.it}
\emailAdd{luigi.tedesco@ba.infn.it}

\abstract
{We study the so-called Gravitational Wave luminosity distance-redshift relation $d_L^{\,GW}(z)$ during cosmological eras driven by non-perfect fluids. In particular, we show that the presence of a shear viscosity in the energy momentum tensor turns out to be the most relevant effect. Within this scenario, a constant shear viscosity imprints the gravitational wave propagation through a friction term $\delta(z)$ with a uniquely given redshift dependence. This peculiar evolution predicts a specific shape for the ratio  $d_{L}^{GW}/d_{L}^{EM}$ which tends to a constant value when the sources are at $z\gtrsim1$, whereas scales linearly with the shear viscosity at lower redshifts, regardless of the value of $\Omega_{m0}$. According to our final discussion, the predicted redshift dependence $\delta(z)$ provided by a shear viscosity could be tested by upcoming surveys of multi-messenger sources against analogous scenarios provided by some widely studied theories of modified gravity.}

\makeatother

\begin{document}
\maketitle \flushbottom

\section{Introduction}
\label{sec:intro}
The detection of Gravitational Waves (GWs) from black holes coalescence \cite{abbott1,abbott2,abbott3,abbott4,abbott5} has opened a new window for exploring new phenomena of interest for astrophysics and cosmology. Moreover, the subsequent detection of GWs emitted by the binary neutron stars \cite{abbott6}, associated to the analysis of the electromagnetic (EM) counterpart \cite{GBM:2017lvd} (in particular $\gamma$-ray burst \cite{Goldstein:2017mmi,Savchenko:2017ffs,Monitor:2017mdv})
led us to the era of the GWs astrophysics and multimessenger
cosmology. In this regards, GWs have opened a new window to understand the
nature of gravity \cite{Deffayet:2007kf,will,yunes,cai,maggiorelibro1,maggiore2}.
Among all these new kinds of observations, the possibility to have an
absolute measurement of the luminosity distance of coalescing compact
binaries has gained an increasing interest \cite{Lombriser:2015sxa}.
Indeed, the signal of coalescing binaries
combines intrinsic and apparent luminosity in order to have a determination of the luminosity distance (from
which the name standard sirens) \cite{dalal,macleod,nissanke,cutler,sathyaprakash,zhao,delpozzo,nishizawa,taylor,camera,tamanini,caprini,cai2,Schutz:1986gp}.

On the other hand, detection of signals like GW170817, observed at
$z\simeq0.009$, can be used to constrain deviations from $\Lambda$CDM model for
what concerns either the nature of gravity or the features of the sources.
One of the most remarkable cases consists of the simultaneous study of electromagnetic
and gravitational signals emitted by the same source: this has provided stringent constraints about
differences in their speed of propagation \cite{Creminelli:2017sry,amendola,Sakstein:2017xjx,Ezquiaga:2017ekz,Baker:2017hug} which, for instance, ruled out entire classes of models such as scalar-tensor and vector-tensor theories of gravity \cite{Creminelli:2017sry,Sakstein:2017xjx,Ezquiaga:2017ekz,Baker:2017hug}.
Another idea to be mentioned has been proposed in \cite{Schutz:1986gp}, where it has been shown that GWs could be used as
an independent tool to study the Hubble-Lema\^itre diagram. In fact,
the localization of the host galaxy for GW170817 allows to
determine $H_{0}$ by the gravitational luminosity distance whereas the electromagnetic counterpart provides the redshift.
Another interesting but futuristic idea regards the possibility to study different messengers from the same source through the so-called time-of-flight distance \cite{Stodolsky:2000aj,Fanizza:2015gdn}.

 The different nature of the gravitational signals can shed light
on the structure and the evolution of the universe up to early epochs
(from the inflationary period to the recombination) that we are unable
to probe directly using EM waves. However, for ``near'' sources,
in the next future we could be able to compare both EM and GW signals.
It is known that distances in astronomy are calibrated using different
techniques, the so-called \textit{distance ladder.} For near astronomical
objects ($z\approx1$), measurements are usually done using the so
called Standard Candles (Supernova IA), from which the accelerated
expansion of the universe was firstly observed in 1998 \cite{key-1}.
It is possible to use also binary mergers as {\it standard sirens},
which could emit GW and also EM when probably one of the objects is
a neutron star. These events are known as Ultranovas and Kilonovas
and in \cite{key-2} it is argued how the latter can be standardized
(like Supernova IA) and how the analysis of the distance given by
the GW measurements (which does not rely on any distance ladder nor
cosmological model), with the redshift, given by the EM counterpart,
might provide new probes for measuring the Hubble constant $H_{0}$.
For the sake of completeness, we mention also the possibility that a GW can be emitted by a source without an EM counterpart. These sources are the so called {\it dark sirens} and their use for the estimation of cosmological parameters is nowadays largely increasing \cite{Leyde:2022fsc}.

In this work we discuss how these new observational probes can be investigated
to detect modifications on the gravitational luminosity
distance as provided by a cosmological viscous fluid.
Indeed, the idea that viscosity might play a role in the cosmological dynamics has been discussed already in the middle 60's \cite{Hawking:1966qi}, where its presence was originally addressed to the gravitational radiation (see also the pioneering works \cite{Misner:1967uu,Matzner1971,Madore:1973xy,Papadopoulos1985}). Recently, the renewed interest in cosmological viscous fluids has increased for different reasons. An interest in the damping of GWs due to the free streaming of neutrinos has been discussed already in the early years of 21st century \cite{Weinberg:2003ur}. However, in this work the focus was in the primordial spectrum of tensor perturbations. In this work, we rather focus on GWs as emitted in the late time Universe.

The most agnostic reason to allow for the presence of viscous fluids is that the perfect fluid description might be a simplistic approximation of the actual Universe. Physically, the emergence of a viscosity might occur in particular scenarios where interaction among the fluids relaxes the thermal equilibrium. According to the specific model in mind, viscosity has been investigated on very early (examining the viscosity effects on the various inflationary observables \cite{Ganguly:2021pke}) and late-time Universe \cite{Brevik:2017msy}. 

Clearly, different kind of viscosity can be considered and the specific nature depends on the particular physical mechanims invoked in the cosmic dynamics. As regards the bulk viscosity, it might play an interesting role in the Universe evolution  \cite{Murphy1973,Arbab1997,Fabris:2005ts,Atreya:2017pny}, since it is compatible with the symmetry requirements of the homogeneous and isotropic FLRW models. From the theoretical point of view, the bulk viscosity may be due to the deviation from the local thermodynamic equilibrium: in this context, the presence of bulk viscosity would lead to an effective pressure in order to restore the thermal equilibrium, broken by the cosmological fluid expanding (or contracting) too much quick. According to this view, hence the effective pressure due to the bulk viscosity would reduce to the "standard" pressure given by the EoS of a perfect fluid as soon as the fluid itself reaches the equilibrium condition.
Also the Dark Energy may be connected with viscosity in the Universe. For instance, \cite{Fabris:2005ts} studies the possibility that the present acceleration of the universe might be driven by a kind of bulk viscous fluid.
However, despite the role of a bulk viscosity might be of interest, along this work we will not consider it, since a bulk viscosity does not lead to an GW attenuation \cite{Prasanna:1999pn,Goswami:2016tsu,key-6}.

On the other hand, we will consider the viscosity in the form of a shear one since it affects directly the evolution of tensor perturbations, leaving the background unaltered. The reason for this choice is that we do not want to study a particular model exhibiting a kind of viscosity but we aim to understand which general features can be understood about the viscous model by analysing the GWs evolution.

For the sake of completeness, we mention that the interest in the shear viscosity occurs in some particular scenarios such as the Self Interacting Dark Matter (SIDM) model. Indeed, the SIDM model is able to generate the shear viscosity in the Universe \cite{Goswami:2016tsu,Atreya:2017pny,Natwariya:2019fif} and explain some features in the small scale structure. More generally, other studies consider Viscous Dark Matter where only shear viscosity is considered (see for example \cite{Blas:2015tla}).

Having this in mind, the machinery that we will adopt to our ends is 
in line with to the one developed for the study of the GWs distance-redshift relation
concerning theories of modified
gravity \cite{key-3,key-4,key-5}. The key aspect that we will show is that a viscous fluid can provide
a non-trivial ratio between the GW and the EM luminosity distance already within the $\Lambda$CDM model.
This interesting aspect emerges when a pure shear viscosity is considered in the cosmic fluid. Indeed, such a modification
of the perfect fluid description leaves unchanged the Friedmann equations but can affect the linear part of the 
perturbations related to the Weyl components of the Riemann tensor, as one would expect 
by a traceless perturbations of the energy-momentum tensor like a shear viscosity.

After having discussed the general framework for the GWs luminosity distance in the above-mentioned scenario,
we will discuss possible observational tests and forecasts for the detection/exclusion of the shear viscosity, according to the specifics of futuristic surveys, such as LISA and Einstein Telescope (ET).

This paper is organized as follows: in Sect. \ref{sec:2}, we recall the evolution
equation for linear gravitational waves in presence of viscous fluids and provide the
general expression for $d_L^{\,GW}(z)$. In Sect. \ref{sec:3}, we first study our results with
a generic power-law solution for the scale factor. Hence we provide an analytical
expression for $d_L^{\,GW}(z)$ in presence of a shear viscosity in
a $\Lambda$CDM-like cosmological model. Possible observational tests and forecasts regarding the presence of
shear viscosity are discussed in Sect. \ref{sec:4}. Conclusions and discussions are finally provided in Sect. \ref{sec:5}.

Conventions: in this
paper we will use the metric signature $g_{\mu\nu}=\text{diag}(+,-,-,-)$;
the Riemann tensor is defined as $R_{\mu\nu\alpha}\,^{\beta}=\partial_{\mu}\Gamma_{\nu\alpha}\,^{\beta}+\Gamma_{\mu\rho}\,^{\beta}\Gamma_{\nu\alpha}\,^{\rho}-\left( \mu\leftrightarrow \nu \right)$,
the Christoffel symbols as $\Gamma_{\mu\nu}\,^{\alpha}=\frac{1}{2}g^{\alpha\beta}(\partial_{\mu}g_{\nu\beta}+\partial_{\nu}g_{\mu\beta}-\partial_{\beta}g_{\mu\nu})$,
the Ricci tensor is given by $R_{\nu\alpha}=R_{\mu\nu\alpha}\,^{\mu}$
and the covariant derivatives act as $\nabla_{\mu}A_{\nu\dots}=\partial_{\mu}A_{\nu}-\Gamma_{\mu\nu}\,^{\alpha}A_{\alpha\dots}+\dots$.

\vspace{1cm}

\section{The Gravitational Wave luminosity distance in viscous scenarios}
\label{sec:2}
Let us suppose to describe the contents of the Universe as a dissipative
fluid. From a physical point of view the energy momentum tensor of such fluid
must depend on the shear viscosity coefficient $\eta_{V}$ and the bulk
viscosity coefficient $\zeta$. By introducing the effective pressure
\begin{equation}
\widetilde{p} =p - \left(\zeta -{2\over 3} \eta_V \right) \nabla_\lambda u^\lambda,
\label{322}
\end{equation}
we have the full energy-momentum tensor of the viscous fluid \cite{thorne2000gravitation,weinberg1972gravitation,Gron:1990ew,rezzolla2013relativistic}
in the following convenient form\footnote{Signs in Eq.~\eqref{323} follow from our metric conventions (see also the derivation provided in \cite{Romatschke:2009im}). Examples of different signatures in Eq.~\eqref{323} due to different metric conventions can be found in \cite{thorne2000gravitation,weinberg1972gravitation,Gron:1990ew,rezzolla2013relativistic}.} (see e.g. \cite{Anand:2017wsj,Montani:2016hmf} for recent discussions)
\begin{equation}
\widetilde{T}_{\alpha\beta} = \left(\rho+ \widetilde{p} \right) u_\alpha u_\beta - \widetilde{p}\,g_{\alpha\beta} - \eta_V \left[ u_\alpha u^\rho \nabla_\rho u_\beta + u_\beta u^\rho \nabla_\rho u_\alpha - \nabla_\alpha u_\beta - \nabla_\beta u_\alpha    \right],
\label{323}
\end{equation}
where $u^{\alpha}$ is the time-like fluid four-velocity ($u_{\alpha}u^{\alpha}=1$).
The above energy momentum tensor
is known to lead to acausal propagation of first order perturbations.
Indeed there exists viable alternatives that have been formulated
which do not possess this anomaly such as the Muller-Israel-Stewart
theories \cite{Muller1967,Israel:1976tn,Israel1979}
that are compatible with our expression for the viscous energy momentum
tensor when the microscopic relaxation time is much smaller than the
time of the macroscopic dissipative process \cite{rezzolla2013relativistic}, so we will assume this hypothesis throughout the work
since we are not considering any particular microphysical model. Moreover
even in MIS theories it is not known if causality is preserved under
general initial conditions beyond the linear regime \cite{Bemfica:2019knx} and generalized theories are being
developed \cite{Bemfica:2019knx,Alqahtani:2017mhy}. Despite the above mentioned critical
aspect of viscous hydrodynamics, it is customary to describe SIDM as a viscous fluid in terms of a first
order gradient expansion as the one we used \cite{Atreya:2017pny,Atreya:2018iom}.
For the purposes of this work, we will only consider the presence of shear viscosity $\eta_{V}$.
This is a simple but interesting case study since only a bulk viscosity can modify the isotropic\footnote{The small anisotropy at the last-scattering surface is typically suppressed in the late-time Universe already if only perfect fluids are considered. When non-perfect fluids are taken into account, relaxing the hypothesis of an isotropic background introduces a coupling between the anisotropy in the metric and the shear viscosity which may indirectly affect the background dynamics. However, in the late-time expanding Universe this anisotropy is exponentially suppressed by the presence of a constant shear viscosity (see Sect. II of \cite{Ganguly:2021pke} for a detailed related discussion).} background equations (see \cite{key-6}).
Hence, neglecting its presence will allow us to study the GWs propagation by using the standard solutions for the background expansion.

Having this in mind, in this section we will analyze what physical consequences
arise from the presence of a shear viscous component in the stress-energy
tensor of the cosmological matter fluid: more precisely, we will study
how the gravitational luminosity distance is affected and how the
gravitational waves propagates in this background. The presence of
a shear viscous source introduces an additional friction term in the
standard gravitational wave equation in an Friedmann-Lema\^itre-Robertson-Walker (FLRW) universe, so we expect
that the overall differences will be in an increased damping effect
of the GW amplitude. We start from the results of Sect. 3.3 of \cite{key-6}. It has been shown there that the perturbed Einstein-Hilbert action and the energy momentum tensor
of a comoving viscous fluid to the second order in the metric, for a conformal
FLRW background metric
\begin{equation*}
g^{FLRW}_{\mu\nu}=a^{2}(\eta)\text{diag}\left( 1,-1,-1,-1 \right)\,,
\end{equation*}
and by making use of the TT gauge, leads to the following propagation
equation for the tensor perturbations
\begin{align}
{h''_i}^{\,j}+2\left(\mathcal{H}+\lambda_{p}^{2}\,\eta_{V}\,a\right)\,{h'_i}^{\,j}-\nabla^{2}{h_i}^j=0\,,
\label{eq:1}
\end{align}
where $\lambda^2_{p}=8\pi G$, $\mathcal{H}=a'/a$, the prime is referred
to derivative with respect to conformal time $\eta$, $\nabla^{2}=\delta^{ij}\partial_{i}\partial_{j}$
and $a(\eta)$ is the FLRW scale factor.

To lighten the notation, let us define $\alpha=\lambda_{p}^{2}\,\eta_{V}$
as the viscosity parameter. We remark that $\alpha$ is a dimensional quantity, scaling as an inverse length.
It is worth then to compare its value to $H_0$, in order to discuss the regimes when shear viscosity could play a role.
For instance, interesting physical values for the shear viscosity have been given in \cite{key-7}, where upper-bounds for $\alpha<10^{-2}\,\text{Mpc}^{-1}\approx 50\,H_0$ have been given by the GW150914 data analyzed supposing a viscous background and by viscous dark matter in galaxies,. To be mentioned is also the upper bound $\alpha<10^{-6}\,\text{Mpc}^{-1}\approx 5\times 10^{-3}\,H_0$, obtained by estimation of the cross section per unit mass mass $\sigma/m$
from Abell 3827 \cite{key-8} in a self-interaction model of dark
matter. For the sake of simplicity, let us then omit the spatial indices
in Eq.~\eqref{eq:1} and
\begin{equation}
h^{\prime\prime}+2\mathcal{H\,}\left[1-\delta(\eta)\right]\,h^{\prime}-\nabla^{2}h=0.\label{eq:2}
\end{equation}
where, following \cite{key-3}, we have defined $\delta(\eta)\equiv -\alpha\,a\,\mathcal{H}^{-1}$.
From Eq.~\eqref{eq:2}, we get that the only effect of a shear viscosity is to alter the evolution equation of a gravitational wave by damping its amplitude through the term $\delta(\eta)$. Since the speed of propagation is not modified, we can follow the same procedure outlined in \cite{key-4} to study the GW luminosity distance, where the authors precisely take into account a modified evolution equation for the GW's which has the same form of Eq.~\eqref{eq:2}. The interesting aspect concerning the viscous model is that the evolution of $\delta$ is fixed. We then recall the main steps in the following.

To this end, it is useful to express $h\left(\eta,\mathbf{x}\right)$ in terms of
its Fourier transform defined as follows\footnote{From the technical point of view, Eq.~\eqref{eq:3} should contain also the polarization tensor $\epsilon_{ij}({\bf k})$ accounting for the two independent polazation modes of $h_i^j$. With any loss of generality, we have omitted it, since Eq.~\eqref{eq:2} provides the same equation for the two polarization states. A massive longitudinal mode can arise in case the GW propagates across molecular medium. A further damping of this mode may occur in particular classes of modified gravity theories \cite{Moretti:2020kpp,Moretti:2021ljj}.}
\begin{equation}
h\left(\eta,\mathbf{\mathbf{x}}\right)=\frac{1}{(2\pi)^{3}}\int d^{3}k\,e^{i\mathbf{k}\cdot\mathbf{x}\,}h_{\mathbf{k}}\left(\eta\right).\label{eq:3}
\end{equation}
From Eq.~\eqref{eq:2} we obtain the time evolution equation for each
mode $k$
\begin{equation}
h_{k}^{\prime\prime}+2\mathcal{H\,}h_{k}^{\prime}+2\alpha\,a\,h_{k}^{\prime}+k^{2}\,h_{k}=0,\label{eq:4}
\end{equation}
where we denote $k=|\mathbf{k}|$ and assume $h_{\mathbf{k}}\left(\eta\right)=h_{k}\left(\eta\right)$.
It is convenient
to rewrite Eq.~\eqref{eq:4} by introducing a new auxiliary variable $v_k(\eta)$ as
\begin{equation}
v_{k}\left(\eta\right)\equiv \widetilde{a}(\eta)\,h_k(\eta),\label{eq:5}
\end{equation}
where we required that the friction term proportional to $v'_k(\eta)$ in the evolution equation for $v_k(\eta)$ is null. Substituting Eq.~\eqref{eq:5} in Eq.~\eqref{eq:4},
we are left with two coupled evolution equations for $v_k$ and $\widetilde{a}$
\begin{equation}
\begin{cases}
v_{k}^{\prime\prime}+\left(k^{2}-\frac{\widetilde{a}^{\prime\prime}}{\widetilde{a}}\right)\,v_{k}=0\\
\frac{\widetilde{a}^{\prime}}{\widetilde{a}}=\mathcal{H}+\alpha\,a\,.
\end{cases}\label{eq:6}
\end{equation}
Interestingly, first of Eq.~\eqref{eq:6} shows that $v_{k}(\eta)$ behaves as
the canonical Mukhanov variable where $\tilde{a}(\eta)$
plays the role of the so-called pump field.
Eqs.~\eqref{eq:6} clearly show
that GWs in presence of shear viscosity couple to an effective scale factor which
fully encodes the viscous content. First of Eqs.~\eqref{eq:6} also shows that on small scales, when
$k^2\gg \tilde{a}''/\tilde{a}$, $v_k$ propagates as free-wave. From Eq.~\eqref{eq:5}, we then conclude that
$h_k$ oscillates as well but its amplitude scales as $\tilde{a}^{-1}$.

On the other hand, general relativistic light-like geodesics in a FLRW geometry are not affected by the presence
of shear viscosity. Hence, the photon wave-vector
scales as $a^{-1}$ \cite{key-3,key-4,key-5,maggiorelibro1}. It follows that we can directly apply
the same results for the GW luminosity-distance relation already discussed for modified gravity theories. We then have
\begin{equation}
\frac{d_{L}^{GW}}{d_{L}^{EM}}\left(z\right)
=\frac{a\left(\eta\right)}{\widetilde{a}\left(\eta\right)},
\label{eq:7}
\end{equation}
where the equality is obtained by imposing $\widetilde{a}\left(\eta_{0}\right)=a\left(\eta_{0}\right)=\widetilde{a}_{0}=a_{0}$
as initial condition in Eq.~\eqref{eq:6}.

Strictly speaking, Eq.~\eqref{eq:7} is fully meaningful when $k^2\gg \tilde{a}''/\tilde{a}$, where we can write
\begin{equation}
\frac{\tilde{a}''}{\tilde{a}}=\frac{a''}{a}+3\,\alpha\,a\mathcal{H}+\alpha^2a^2\,.
\end{equation}
Moreover, the GWs that we can actually probe propagates on sub-horizon scales. This condition gives $k\eta\gg 1$ for a class of cosmological models such as power-law scale factor or $\Lambda$CDM, where it also holds $a''/a\sim \eta^{-2}$ and $\mathcal{H}\sim \eta^{-1}$. This means that the requirement $k^2\gg \tilde{a}''/\tilde{a}$ translates to
\begin{equation}
\left(\frac{k}{\mathcal{H}}\right)^2\gg 1+3\,a\,\frac{\alpha}{\mathcal{H}}+a^2\,\left(\frac{\alpha}{\mathcal{H}}\right)^2\,.
\label{eq:condition}
\end{equation}
Since $a<1$, for the redshift of interest for upcoming observations such as LISA \cite{LISA:2017pwj,LISACosmologyWorkingGroup:2022jok} and ET \cite{Sathyaprakash:2012jk}, i.e. $z\lesssim 2$, Eq.~\eqref{eq:condition} tells us that sub-horizon modes can be safely considered for our models as long as $\alpha\,H^{-1}_0 \lesssim 1$. Instead, for higher values of $\alpha$ our results are meaningful for $k\gg\alpha$. It is worth to remark that when $\alpha\gtrsim H_0$, not all the subhorizon modes are compatible with our discussion, since the scale introduced by $\alpha$ lays within the horizon itself. However, our analysis is still viable for the range of wavelengths compatible with the current and forthcoming interferometers. Indeed, for the case of quite large shear viscosity $\alpha = 10 H_0$, we would have that the spatial scale of $\alpha^{-1}$ is roughly 10 times smaller than the horizon. This scale is anyway much larger than the wavelength range that Earth and Space based interferometers are able to probe, ensuring then that the condition $k\gg\alpha$ is always safely satisfied for the modes of our interest.

In the following sections we will discuss the explicit cases for
a power-law scale factor and for the $\Lambda$CDM model. 

\vspace{1cm}

\section{Viscous models}
\label{sec:3}
In this section, we study the GW distance-redshift relation for some particular cases of physical interest. Firstly, we analyze the case of a generic power law behaviour for the scale factor. An interesting application for this case could provided by models where also the bulk viscosity is not null. Indeed, in this case the Friedmann equations are modified by the bulk viscosity and some interesting deviation from the Cosmic Concordance model might occur. Secondly, we will discuss the case of a viscous extension of $\Lambda$CDM model, where only shear viscosity is allowed.

\subsection{Power-law solution}
We first present the illustrative case of a power-law solution for the cosmological background.
Under these assumptions, the scale factor in term of the conformal time $\eta$ reads
\begin{equation}
a\left(\eta\right)=a_{0}\left(\frac{\eta}{\eta_{0}}\right)^{\beta},\label{eq:8}
\end{equation}
and this yields to $\mathcal{H}=\frac{\beta}{\eta}$, where $\eta_{0}$ is the conformal time today.
Inserting the last two equalities in the second line of Eq.~\eqref{eq:6}
we then have
\begin{equation}
\frac{\widetilde{a}^{\prime}}{\widetilde{a}}=\frac{\beta}{\eta}+\alpha\,a_{0}\left(\frac{\eta}{\eta_{0}}\right)^{\beta}\label{eq:10}
\end{equation}
with the initial condition given by $\widetilde{a}\left(\eta_{0}\right)=a\left(\eta_{0}\right)$.
We have done this choice in order to recover $d_L^{\,GW}(0)=d_L^{\,EM}(0)$ (see Eq.~\eqref{eq:7}. For $\beta\neq-1$, we then have
\begin{equation}
\widetilde{a}\left(\eta\right)=a(\eta)\exp\left[\frac{\alpha\,a_{0}\left(\eta^{\beta+1}-\eta_{0}^{\beta+1}\right)}{(\beta+1)\eta_{0}^{\beta}}\right].\label{eq:13}
\end{equation}
A direct comparison between Eqs.~\eqref{eq:8} and \eqref{eq:13} shows that the damping
effects introduced by the shear viscosity is an exponential suppression in the past of the effective
scale factor $\tilde{a}$, which governs the GWs propagation.

\begin{figure}[ht!]
\begin{centering}
\includegraphics[scale=1]{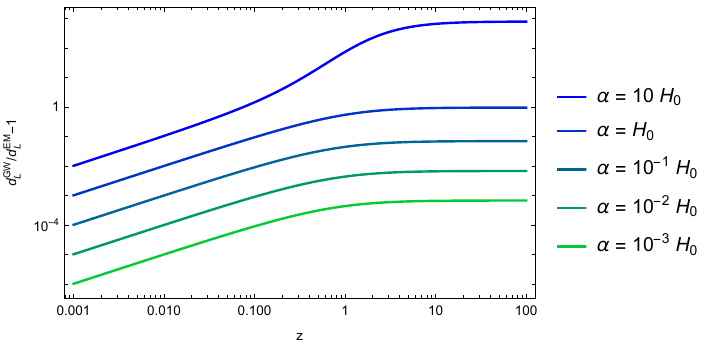} 
\par\end{centering}
\caption{Luminosity distance ratio for different values of $\alpha$ in unit of $H_0$
for CDM model.}
\label{fig:Luminosity-distance-ratio}
\end{figure}

Finally, we substitute Eq.~\eqref{eq:13} in Eq.~\eqref{eq:7} to
obtain
\begin{equation}
\frac{d_{L}^{GW}\left(\eta\right)}{d_{L}^{EM}\left(\eta\right)}
=\exp\left\{\frac{\alpha\,a_{0}\eta_0}{(\beta+1)}\left[1-\left(\frac{a}{a_0}\right)^{(\beta+1)/\beta}\right]\right\}\,.
\label{eq:14}
\end{equation}
Since our observations relate distances to redshift, the last equation
\eqref{eq:14} can be further manipulated in order to make explicit
the dependence on $z$. To this purpose, we first recall that $1+z=\frac{a_{0}}{a}$.
Hence by expressing the conformal time today in terms of the Hubble constant $H_0$ we obtain
\begin{equation}
\frac{d_{L}^{GW}}{d_{L}^{EM}}\left(z\right)
=\text{exp}\left[\frac{\beta}{\beta+1}\frac{\alpha}{H_0}\left(1-\frac{1}{\left(1+z\right)^{\frac{\beta+1}{\beta}}}\right)\right]\,.
\label{eq:17}
\end{equation}
This is the most general equation that relates the gravitational luminosity
distance to the redshift in a power law Universe with constant shear
viscosity. Setting $\beta=2$, we apply the general result given in
Eq.~\eqref{eq:17} to the physical scenario of a matter dominated Universe
(Einstein-de Sitter Universe). In Fig.~\ref{fig:Luminosity-distance-ratio} we
show the relative correction between the gravitational and the 
electromagnetic luminosity distance-redshift relation as given by Eq.~\eqref{eq:17} for different values of viscosities. 

Regardless of the value of $\alpha$, we notice that all the solutions tend to a constant value of the ratio
beyond a certain redshift $z^*$. For a generic value of $\beta$, this overall offset is given by
\begin{equation}
\frac{d_{L}^{GW}}{d_{L}^{EM}}\approx e^{\frac{\alpha}{H_0}\frac{\beta}{(\beta+1)}}\,,
\label{eq:18}
\end{equation}
which clearly shows that is controlled by the ratio $\alpha/H_0$.
Before moving on, a comment about the behavior for large $\alpha/H_0$ in Fig.~\ref{fig:Luminosity-distance-ratio} is in order. The intermediate bump emerging when $\alpha = 10\,H_0$ is due to the nonlinear modification of $\tilde{a}$ with respect to $a$, as accounted for in the second of Eqs.~\eqref{eq:6}. With the proviso that $k^2\gg\tilde{a}''/\tilde{a}$, $h_k$ scales as $\tilde{a}^{-1}$, regardless of $k$. Hence we get that the scaling of the GW amplitude is entirely given by Eq.~\eqref{eq:17}, as a consequence of the second of Eqs.~\eqref{eq:6}. When $\alpha\ll H_0$, Eq.~\eqref{eq:17} gives
\begin{equation}
\frac{d^{GW}_L}{d^{EM}_L}(z)=1
+\frac{\alpha}{H_0}\frac{\beta}{\beta+1}\left[1-\left(1+z\right)^{-\frac{\beta+1}{\beta}}\right]
+\mathcal{O}\left( \left(\frac{\alpha}{H_0}\right)^2 \right)\,,
\end{equation}
which only exhibits two regimes when $z$ is compared to 1. However, when $\alpha \gtrsim H_0$, the nonlinear dependence of $d^{GW}_L/d^{EM}_L$ cannot be neglected. In particular, since both small ($z\ll 1$) and large ($z\gg 1$) redshift behaviors are universal in regard to the value of $\alpha H^{-1}_0$, the intermediate range at $z\sim 1$ is precisely the only range where the nonlinear dependence on the shear viscosity can change the qualitative form of Fig.~\ref{fig:Luminosity-distance-ratio}.

Although an Einstein-de Sitter universe is a very good approximation of the
evolution of the Universe after the CMB, it fails to reproduce the
recent Dark Energy epoch and a better model of the Universe is required.
In the next subsection we will discuss the effects of viscosity within
a $\Lambda$CDM-like model.

\subsection{A viscous $\Lambda$CDM model}

After recombination, the radiation contribution to the expansion of
the Universe has become negligible.
Ignoring the radiation component may underestimate observables related to the primordial tensor perturbations (such as 2-point correlation of the lensing shear \cite{Fanizza:2022wob}), especially on scales small enough to have re-entered the horizon before the recombination. However, the GW's analysed with the $d^{GW}_L(z)$ are not related to the primordial tensor spectrum, as they are emitted only in recent epochs (up to $z\sim 5$), where the radiation component can be safely neglected.

The first difficulty we have to deal with is the absence of an analytic expression for $a(\eta)$
in the Cosmic Concordance model. In order to circumvent this technical
aspect, it is convenient to rewrite the second line of Eq.~\eqref{eq:6}
in terms of the redshift. With this in mind, we first provide the
following well-known relations ($a_{0}=1$)
\begin{equation}
\frac{d}{d\eta}=-H(z)\frac{d}{dz}\qquad\text{and}\quad
\mathcal{H}(z)=\frac{H(z)}{1+z}\,,
\label{eq:19}
\end{equation}
where
\begin{equation}
H(z)=H_{0}\sqrt{\Omega_{m0}(1+z)^{3}+\Omega_{\Lambda0}}\,,\label{eq:23}
\end{equation}
and clearly $\Omega_{\Lambda0}+\Omega_{m0}=1$. Therefore, first of Eqs.~\eqref{eq:6} becomes\footnote{Here a possible time dependence in $\alpha$ is considered for a sake of generality. This case will not be considered along this work but might be of interest for other scenarios, where it has been claimed that a running value $\alpha$ could emerge from the coarse graining of the small-scales inhomogeneities \cite{Blas:2015tla,Floerchinger:2016hja}.}
\begin{equation}
\frac{d\ln\widetilde{a}}{dz}=-\frac{1}{1+z}-\frac{\alpha(z)}{H(z)(1+z)}\,.
\label{eq:20}
\end{equation}
By integrating the lhs between $\widetilde{a}_{0}=a_{0}=1$ and $\widetilde{a}$
and the rhs between 0 and $z$ we obtain
\begin{equation}
\widetilde{a}(z)=\frac{1}{1+z}\exp\left(-\int_{0}^{z}\frac{\alpha(z')}{(1+z')H(z')}dz'\right)\,,\label{eq:21}
\end{equation}
Once inserted in Eq.~\eqref{eq:7}, Eq.~\eqref{eq:21} gives 
\begin{equation}
\frac{d_{L}^{GW}}{d_{L}^{EM}}\left(z\right)=\text{exp}\left[\int_{0}^{z}\frac{\alpha(z')}{(1+z')H(z')}dz'\right].\label{eq:22}
\end{equation}
Last equation is a general expression that holds true in any FLRW
cosmology with a viscous fluid component in the stress-energy tensor,
where the shear viscosity parameter is not constant. All the information
about the background gravitational sources is encoded in $H(z)$.

We remark that Eq.~\eqref{eq:22} holds true also for values of $H(z)$ which are more general of Eq.~\eqref{eq:23}.
Indeed, as long as the Hubble function is evaluated accordingly to the total
stress-energy tensor, it follows that the equation is valid even if
the fluids do not have a barotropic equation of state (e.g. polytropic
fluids) and it remains valid also in the presence of bulk viscous
contributions. However, since now on, we will only consider background dynamics
driven by dark matter and $\Lambda$, in agreement with the value of $H(z)$ provided in
Eq.~\eqref{eq:23}.

With this in mind, we then combine Eq.~\eqref{eq:22} and with Eq.~\eqref{eq:23}
and consider a constant $\alpha(z)=\alpha$. We underline that the
shear viscous parameter $\alpha$ in a model with a mixture of viscous
fluids is given by $\alpha=\sum_{A}\alpha_{A}$, where $A$ is an
index that runs over the different viscous fluids.
As a consequence, with this condition, in a $\Lambda$CDM-like model it is not possible to distinguish which fluid
component between Dark Energy, Dark Matter
or both is responsible for the shear viscous effects on the gravitational
wave propagation. However, if one has in mind a minimal extension to the $\Lambda$CDM
model, the viscous property should be addressed only to the Dark Matter fluid.
This is because a Dark Energy as given by a cosmological constant is
the only term compatible with General Relativity, according to the hypothesis of the Lovelock theorem.
This minimal description admits then an EoS for $\Lambda$ which is nothing but the one of a perfect fluid.
The integral in Eq.~\eqref{eq:22} hence becomes
\begin{equation}
\int_{0}^{z}\frac{1}{(1+z')H(z')}dz'=\frac{1}{3H_{0}\sqrt{\Omega_{\Lambda0}}}\ln\left(\frac{H(z)/H_0-\sqrt{\Omega_{\Lambda0}}}{H(z)/H_{0}+\sqrt{\Omega_{\Lambda0}}}\,\frac{1+\sqrt{\Omega_{\Lambda0}}}{1-\sqrt{\Omega_{\Lambda0}}}\right)\label{eq:24}
\end{equation}
bringing then to
\begin{equation}
\frac{d_{L}^{GW}}{d_{L}^{EM}}\left(z\right)=\left(\frac{\sqrt{\Omega_{m0}(1+z)^{3}+\Omega_{\Lambda0}}-\sqrt{\Omega_{\Lambda0}}}{\sqrt{\Omega_{m0}(1+z)^{3}+\Omega_{\Lambda0}}+\sqrt{\Omega_{\Lambda0}}}\,\frac{1+\sqrt{\Omega_{\Lambda0}}}{1-\sqrt{\Omega_{\Lambda0}}}\right)^{\frac{\alpha}{3H_{0}\sqrt{\Omega_{\Lambda0}}}}\,.
\label{eq:25}
\end{equation}

\begin{figure}[ht!]
\begin{centering}
\includegraphics[scale=0.8]{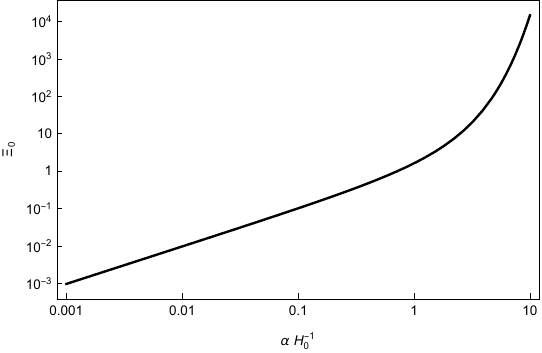}
\includegraphics[scale=0.8]{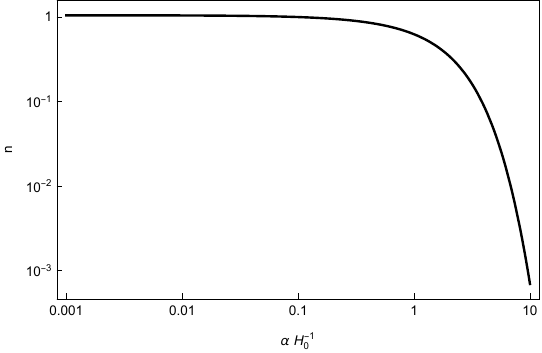}
\caption{Dependence of $\Xi_0$ (left) and $n$ (right) in terms of $\alpha H^{-1}_0$, according to the widely used parametrization given in Eq.~\eqref{eq:parametrization}. The value of $\Omega_{m0}=0.307$ has been taken in accordance to the local estimation from the Pantheon sample \cite{Pan-STARRS1:2017jku}.}
\label{fig:parameters}
\par\end{centering}
\end{figure}

Just as discussed in the previous section, we have again a critical redshift $z^*$ such that, when $z>z^*$, we have:
\begin{equation}
\Xi_0\equiv\frac{d_{L}^{GW}}{d_{L}^{EM}}(z\gg z^*)=\left(\frac{1+\sqrt{\Omega_{\Lambda0}}}{1-\sqrt{\Omega_{\Lambda0}}}\right)^{\frac{\alpha}{3H_{0}\sqrt{\Omega_{\Lambda0}}}}\,.
\label{eq:25a}
\end{equation}
Here we have called this limit value $\Xi_0$ since we have in mind the widely used parametrization of $d_L^{\,GW}/d_L^{\,EM}$ proposed in \cite{key-4}
\begin{equation}
\frac{d_L^{\,GW}}{d_L^{\,EM}}(z)=\Xi_0 + \frac{1-\Xi_0}{(1+z)^n}\,.
\label{eq:parametrization}
\end{equation}
According to Eq.~\eqref{eq:25}, the parametrization in Eq.~\eqref{eq:parametrization} is not expected to work perfectly for any value of $\alpha$.
\begin{figure}[ht!]
\begin{centering}
\includegraphics[scale=1]{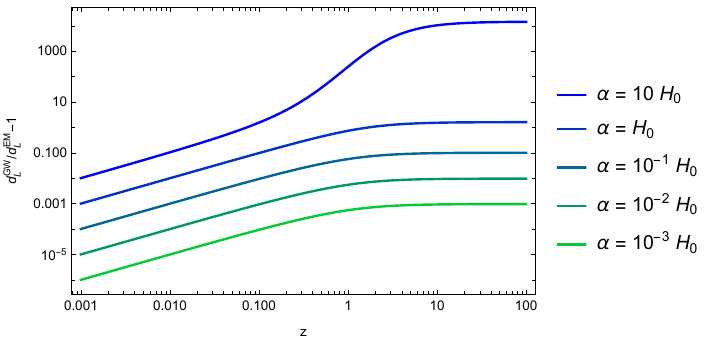}
\par\end{centering}
\caption{Luminosity distance ratio for different values of $\alpha$ in unit of $H_0$
for $\Lambda$CDM where $\Omega_{m0}=0.307$ has been taken in accordance to the local estimation from the Pantheon sample \cite{Pan-STARRS1:2017jku}.}
\label{fig:Luminosity-distance-ratio-LCDM}
\end{figure}
Indeed, as we can see from Fig.~\ref{fig:Luminosity-distance-ratio-LCDM}, for high value of $\alpha/H_0$, the emergence of an intermediate slope at redshifts $\sim 1$ renders Eq.~\eqref{eq:parametrization} hard to be used. However, when $\alpha$ is perturbatively smaller than $H_0$, Eq.~\eqref{eq:parametrization} works quite well.
\begin{figure}[ht!]
\begin{centering}
\includegraphics[scale=1]{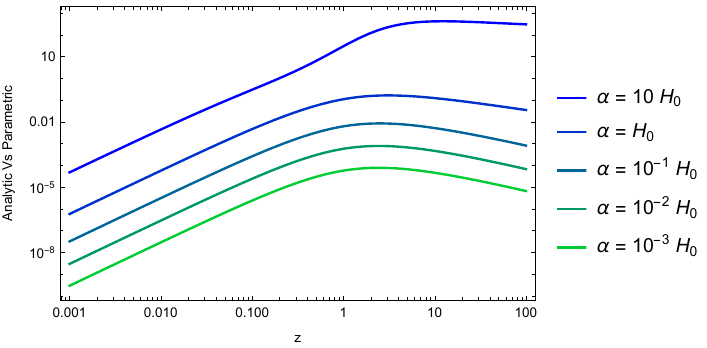}
\caption{Relative error between the exact expression in Eq.~\eqref{eq:25} and the parametric expression in Eq.~\eqref{eq:parametrization}. We fix $\Omega_{m0}=0.307$ in accordance to the local estimation of the Pantheon sample \cite{Pan-STARRS1:2017jku}. This plot shows that the widely used parametrization in Eq.~\eqref{eq:parametrization} is reliable only for small values of $\alpha\,H^{-1}_0$.}
\label{fig:anVSpar}
\par\end{centering}
\end{figure}
This is more evident in Fig.~\ref{fig:anVSpar}, where the slope $n$ in Eq.~\eqref{eq:parametrization} has been modeled as (see \cite{LISACosmologyWorkingGroup:2019mwx} for a detailed discussion about different parametrization of $n$)
\begin{equation}
n\equiv\frac{\delta(0)}{1-\Xi_0}\,.
\end{equation}
The dependences of $\Xi_0$ and $n$ on the ratio $\alpha/H_0$ are shown in Fig.~\ref{fig:parameters}, where $\Omega_{m0}=0.307$ has been taken in accordance to \cite{Pan-STARRS1:2017jku}.
From both these panels, we notice that the discrepancy between the gravitational and electromagnetic
luminosity distance deviates exponentially when the shear viscosity becomes of order of $H_0$.

What emerges from Figs.~\ref{fig:anVSpar} and \ref{fig:parameters} is that the parametrization of Eq.~\eqref{eq:parametrization} can be safely adopted also for the viscous models when its value is at maximum of few percent of the $H_0$. Accordingly, also the constraints on the value of $\alpha$ from Eq.~\eqref{eq:parametrization} are not trustworthy beyond this regime. The reason for this discrepancy stands in the fact that Eq.~\eqref{eq:parametrization} holds when $d_L^{\,GW}/d_L^{\,EM}$ exhibits only two regimes: top curve of Fig.~\ref{fig:Luminosity-distance-ratio-LCDM} clearly shows that this is not the case when $\alpha\gg H_0$. We remark that in this case the estimation of $n$ becomes meaningless whereas $\Xi_0$ can still be interpreted as the limit value of $d_L^{\,GW}/d_L^{\,EM}$.

As a final remark we show that in the limit $\Omega_{\Lambda0}\rightarrow0$
we retrieve the Einstein-de Sitter solution given in Eq.~\eqref{eq:17}
by setting $\beta=2$. To this end, it is more straightforward
to expand in terms of $\Omega_{\Lambda 0}$ Eq.~\eqref{eq:24}, having in mind the constraint
$\Omega_{m0}=1-\Omega_{\Lambda0}$. Since we are interested in the leading term and the
logarithm is divided $\sqrt{\Omega_{\Lambda0}}$, we neglect terms of order
$\mathcal{O}(\Omega_{\Lambda0})$ or higher.
First fraction in the argument of the logarithm
of Eq.~\eqref{eq:24} then becomes
\begin{align}
\frac{\sqrt{\Omega_{m0}(1+z)^{3}+\Omega_{\Lambda0}}-\sqrt{\Omega_{\Lambda0}}}{\sqrt{\Omega_{m0}(1+z)^{3}+\Omega_{\Lambda0}}+\sqrt{\Omega_{\Lambda0}}}
=&\frac{\sqrt{1-\Omega_{\Lambda0}+\Omega_{\Lambda0}(1+z)^{-3}}-\sqrt{\Omega_{\Lambda0}}(1+z)^{-3/2}}{\sqrt{1-\Omega_{\Lambda0}+\Omega_{\Lambda0}(1+z)^{-3}}+\sqrt{\Omega_{\Lambda0}}(1+z)^{-3/2}}
\nonumber\\
=&\frac{1-\sqrt{\Omega_{\Lambda0}}(1+z)^{-3/2}}{1+\sqrt{\Omega_{\Lambda0}}(1+z)^{-3/2}}+\mathcal{O}\left( \Omega_{\Lambda0} \right)
\nonumber\\
=&1-2\,\sqrt{\Omega_{\Lambda0}}\,(1+z)^{-3/2}+\mathcal{O}\left( \Omega_{\Lambda0} \right)\,.
\label{eq:26}
\end{align}
In the same way, for the second fraction in Eq.~\eqref{eq:24} we have
\begin{equation}
\frac{1+\sqrt{\Omega_{\Lambda0}}}{1-\sqrt{\Omega_{\Lambda0}}}
=1+2\sqrt{\Omega_{\Lambda0}}+\mathcal{O}(\Omega_{\Lambda0})\,.\label{eq:30}
\end{equation}
Hence, thanks to Eqs.~\eqref{eq:26} and \eqref{eq:30}, the expansion of the logarithm in Eq.~\eqref{eq:24} is
\begin{equation}
\ln\left(\frac{H(z)/H_0-\sqrt{\Omega_{\Lambda0}}}{H(z)/H_{0}+\sqrt{\Omega_{\Lambda0}}}\,\frac{1+\sqrt{\Omega_{\Lambda0}}}{1-\sqrt{\Omega_{\Lambda0}}}\right)
=2\sqrt{\Omega_{\Lambda0}}\left[1-(1+z)^{-3/2}\right]+\mathcal{O}\left(\Omega^{3/2}_{\Lambda0}\right)\,.
\label{eq:31}
\end{equation}
By inserting this last expansion in Eq.~\eqref{eq:24}, we then arrive
at
\begin{equation}
\int_{0}^{z}\frac{1}{(1+z')H(z')}dz'=\frac{2}{3H_{0}}\left[1-(1+z)^{-3/2}\right]+\mathcal{O}\left(\Omega_{\Lambda0}\right)\,.
\label{eq:32}
\end{equation}
Finally, by using Eq.~\eqref{eq:22} with a constant $\alpha$, we recover the result of Eq.~\eqref{eq:17}.

\vspace{1cm}

\section{A smoking gun for the shear viscosity?}
\label{sec:4}
An interesting feature emerging from the discussion in the previous section is the neat
distinction between two regimes of the ratio $d_L^{\,GW}/d_L^{\,EM}$ for small values of
$\alpha/H_0$. The crossing from one regime to the other occurs at a given redshift, where
a knee between the linear and constant behaviour is evident. To quantify this transition,
we first expand Eq.~\eqref{eq:25} for low redshift
\begin{equation}
\frac{d_L^{\,GW}}{d_L^{\,EM}}=1-\delta(0)\,z +\mathcal{O}\left( z^2 \right)=1+\frac{\alpha}{H_0} z +\mathcal{O}\left( z^2 \right)\,,
\label{eq:linear}
\end{equation}
where, following \cite{key-3}, we have used $\delta(0)=-\alpha\,H^{-1}_0$.
It is already interesting to notice from Eq.~\eqref{eq:linear} that the linear regime appearing at low redshift of
Fig.~\ref{fig:Luminosity-distance-ratio-LCDM} is independent of $\Omega_{m0}$:
a detection of a linear increasing behaviour for close sources is entirely addressed to a
non-null shear viscosity in this scenario, regardless of the value of $\alpha$.
However, the amplitude still look prohibitive according to the forecasted precision at those redshifts.
Indeed, following \cite{key-4}, we can use the
the stringent limit on $\delta(0)$ coming from combined observations of the standard siren
GW170817 \cite{abbott6} and electromagnetic counterpart of its host galaxy NGC4993 \cite{Freedman:2010xv,Cantiello:2018ffy}. This provides the bounds $\delta(0)= -7.8^{+9.7}_{-18.4}$. Despite this is a test to directly constrain $\alpha H^{-1}_0$,
the provided bound is unfortunately not so stringent.

Stronger constraints can be put by considering the forecasts for LISA \cite{LISACosmologyWorkingGroup:2019mwx} and ET \cite{key-4}. To this end, we make use of the parametrization in Eq.~\eqref{eq:parametrization} and consider the forecasted error $\Delta\Xi_0$ respectively given by $0.044$ for LISA \cite{LISACosmologyWorkingGroup:2019mwx} and $0.008$ for ET \cite{key-4}. Indeed, for small values of $\alpha H^{-1}_0$, Eq.~\eqref{eq:25a} gives
\begin{equation}
\Xi_0\approx 1+\frac{1}{3}\frac{\alpha}{H_0}\ln\left( \frac{1+\sqrt{\Omega_{\Lambda 0}}}{1-\sqrt{\Omega_{\Lambda 0}}} \right)\,.
\label{eq:42}
\end{equation}
Eq.~\eqref{eq:42} we can then be inverted to get the ratio $\alpha/H_0$ as a function of the parameters $\Omega_{m0}$ and $\Xi_0$. In this way, the precision on the estimation of $\alpha/H_0$ propagating from the error bars of $\Xi_0$ and $\Omega_{m0}$ is
\begin{equation}
\Delta\left( \frac{\alpha}{H_0} \right)
=\frac{3\,\Delta\Xi_0}{\ln\left( \frac{1+\sqrt{\Omega_{\Lambda 0}}}{1-\sqrt{\Omega_{\Lambda 0}}}  \right)}
+\frac{3\,|1-\Xi_0|\,\Delta\Omega_{m0}}{\sqrt{\Omega_{\Lambda0}}\,\Omega_{m 0}\ln^2\left( \frac{1+\sqrt{\Omega_{\Lambda 0}}}{1-\sqrt{\Omega_{\Lambda 0}}}\right)}\,.
\end{equation}
Around the fiducial values of the standard cosmic concordance model ($\Xi_0=1$), we then have that $\alpha/H_0$ could be constrained by future missions with an error given by
\begin{equation}
\Delta\left( \frac{\alpha}{H_0} \right)\approx\frac{3\,\Delta\Xi_0}{\ln\left( \frac{1+\sqrt{\Omega_{\Lambda 0}}}{1-\sqrt{\Omega_{\Lambda 0}}}  \right)}
\approx\Delta\Xi_0\,,
\label{eq:forecasts}
\end{equation}
where we have used the value $\Omega_{m0}=0.307\pm0.012$ as given by the independent estimation from Pantheon sample \cite{Pan-STARRS1:2017jku}. We remark that Eq.~\eqref{eq:forecasts} follows from the optimistic case $\Xi_0=1$ where the contribution of $\Delta\Omega_{m0}$ to the ultimate forecasts cancels out. However, our result is reliable also for more conservative scenarios. Indeed, if we let $\Xi_0$ deviate from 1 by the same order of magnitude permitted by the error bars (namely $|1-\Xi_0|\sim \Delta\Xi_0$), we get that our forecast in Eq.~\eqref{eq:forecasts} is worsen only by $\sim 1\%$, according to our chosen cosmological parameters.

Interestingly, the forecasted error $\alpha/H_0$ is almost of the same order of magnitude as the one for $\Xi_0$. This opens a new window to investigate whether a shear viscosity of order of few percents of $H_0$ is present in the cosmological fluids. We also remark that these constraints are more stringent than the ones provided by the same analysis for particular classes of modified gravity theories, where the forecasted errors for the free parameters are worsen by almost one order of magnitude wrt to $\Delta\Xi_0$ \cite{Frusciante:2021sio}.

As a further remark, an interesting idea would be to use Eq.~\eqref{eq:forecasts} as a forecast for $\alpha$ itself, rather than just $\alpha\,H^{-1}_0$. This could be achieved by combining Eq.~\eqref{eq:forecasts} with the analysis of the expected Hubble-Lema\^itre diagram from Superluminous Supernovae \cite{DES:2020faa} for forthcoming Large-Scale Structure surveys at higher redshift (up to $z=4$). Indeed, it has been shown in \cite{Fanizza:2021tuh} that this analysis could provide a (model dependent) estimation of $H_0$ at a precision level of $\sim0.1\%$.

Finally, we discuss another test that could be done by looking at the transition between the linear evolution and
the constant offset given by Eq.~\eqref{eq:25a}. This transition would occur at the redshift $z^*$
when the linear growth in Eq.~\eqref{eq:linear} reaches the offset value in Eq.~\eqref{eq:25a}.
This simple reasoning returns an expression for $z^*$ given by
\begin{equation}
z^*= \frac{H_0}{\alpha}\left[\left(\frac{1+\sqrt{1-\Omega_{m0}}}{1-\sqrt{1-\Omega_{m0}}}\right)^{\frac{\alpha}{3H_{0}\sqrt{1-\Omega_{m0}}}}-1\right]\,.
\label{eq:zstar}
\end{equation}
We remark that Eq.~\eqref{eq:zstar} is meaningful only for small values of the ratio $\alpha/H_0$, since the transition
between the linear and the constant regime occurs directly only in this case.
Indeed, Fig.~\ref{fig:Luminosity-distance-ratio-LCDM} for values
$\alpha/H_0 >1$, a third intermediate regime appears and this breaks the validity of the discussed case study.
Fortunately, the values of $\alpha$ when this limitation occurs are too high to be seriously considered.
It is worth to notice that when $\alpha\ll H_0$, within the uncertainties of $\Omega_{m0}=0.307\pm 0.012$ given by \cite{Pan-STARRS1:2017jku}, the value of $z^*$ is quite stable around the value of $z^*=1$.

\vspace{1cm}

\section{Conclusions and outlook}
\label{sec:5}
In this paper we have studied the influence of shear viscosity on
the so-called gravitational luminosity distance. The main property of the shear viscosity 
is that it leaves unchanged the Friedmann equations and modifies the linear tensor perturbations
by an additional friction terms. As a consequence, the most stringent constraints on the speed
of GWs propagation are satisfied but a lot of room for other effects can be investigated.

In particular, the friction term $\delta(\eta)$ arising from the viscosity impacts the GW luminosity distance-redshift relation
in two specific ways:
\begin{itemize}
\item{the shear viscosity introduces a new length scale $\alpha$;}
\item{the friction term depends on the Hubble function through the ratio $\alpha/\mathcal{H}$.}
\end{itemize}
From these two properties we infer that the relevant parameter to study and/or discard the presence of the shear viscosity in the cosmological fluid is the relative amplitude between $\alpha$ and $H_0$. In this regard, we have shown that when $\alpha H^{-1}_0\ll 1$, most of the well known techniques for the study of GW luminosity distance in modified gravity theories can be directly applied.

More specifically, for a $\Lambda$CDM-like viscous model, the gravitational distance given in Eq.~\eqref{eq:25} exhibits a constant asymptotic value at redshifts $z>1$ as given by the parameter $\Xi_0$. For the case of perturbative amplitude of $\alpha/H_0$, $\Xi_0$ can actually be interpreted as usually done \cite{key-4} for the parametrization of Eq.~\eqref{eq:parametrization} and consequently forecast whether it will be possible to discriminate the presence shear viscosity from futuristic surveys, such as LISA and ET. Our analysis indicates that non-perfect fluids with a shear viscosity could be studied in details up to values of the $\alpha$ of few percents of $H_0$.

On the contrary, non-perturbative values of the shear viscosity ($\alpha\gtrsim H_0$) could be easily distinguished from all the class of models compatible with Eq.~\eqref{eq:parametrization} and there are two reasons for that:
\begin{itemize}
\item{first of all, the offset value $\Xi_0$ exponentially grows with $\alpha/H_0$. This means that a large deviation of $d^{\,GW}_L$ from the EM luminosity distance could be easily detectable, if present.}
\item{Secondly, the redshift dependence of $d_L^{\,GW}/d_L^{\,EM}$ largely differs from the parametric fitting functions in Eq.~\eqref{eq:parametrization}, also at intermediate redshifts around $z\sim 1$ (see the curve for $\alpha = 10\,H_0$ in Fig.~\ref{fig:Luminosity-distance-ratio-LCDM}).}
\end{itemize}
As a final remark, we stress that the behaviour of $d_L^{\,GW}/d_L^{\,EM}$ for small redshifts ($z\ll 1$), is insensitive to $\Omega_{m0}$ and depends only linearly on the ratio $\alpha/H_0$, regardless of its amplitude. For this kind of close sources, the independent estimation of $z$, $d_L^{\,EM}$ and $d_L^{\,GW}$ may allow to have independent measurements of $H_0$ and $\alpha$ rather than just their ratio. This is also likely in view of the use the three-dimensional cross-correlation technique for GWs sources and galaxies \cite{Mukherjee:2020hyn,Mukherjee:2020mha}.
Such a scenario could provide a highly promising test to estimate the shear viscosity if the precision concerning the measurement of $d_L^{\,GW}$ would dramatically increase in the next years.

\vspace{1cm}

\section*{Acknowledgements}
The authors wish to warmly thank Maurizio Gasperini for his comments and suggestions.
GF is thankful to Jos� Fonseca and Noemi Frusciante for valuable comments on the results presented in this work. 
GF acknowledges support by Funda\c{c}\~{a}o para a Ci\^{e}ncia e a Tecnologia (FCT) under the program {\it ``Stimulus"} with the grant no. CEECIND/04399/2017/CP1387/CT0026 and through the research project with ref. number PTDC/FIS-AST/0054/2021. GF is also member of the Gruppo Nazionale per la Fisica Matematica (GNFM) of the Istituto Nazionale di Alta Matematica (INdAM). LT is supported in part by INFN under the program TAsP ({\it ``Theoretical Astroparticle Physics"}), and by the research grant number 2017W4HA7S {\it ``NAT-NET: Neutrino and Astroparticle Theory Network"}, under the program PRIN 2017 funded by the Italian Ministero dell'Universit\`a e della Ricerca (MUR).

\vspace{1cm}

\bibliographystyle{JHEP}
\bibliography{biblio}

\end{document}